\documentstyle[twocolumn,aps,epsf]{revtex}

\def\y{\'{\i}}
\def\ni{\noindent}
\def\beq{\begin{equation}}
\def\eeq{\end{equation}}
\input{psfig}

\begin{document}
\title{Low Energy Pion-Hyperon Interaction}
\author{C.C. Barros Jr. and Y. Hama}
\address{Instituto de F\y sica, Universidade de S\~ao Paulo\\
C.P. 66318, 05315-970 S\~ao Paulo-SP, Brasil}
\maketitle

\begin{abstract}
We study the low energy pion-hyperon interaction considering effective
non-linear chiral invariant Lagrangians including pions, $\rho$-mesons, 
$\sigma$-mesons, hyperons and corresponding resonances. Then we calculate 
the $S$- and $P$-wave phase-shifts, total cross sections, angular distributions and polarizations for the momentum in the center-of-mass frame up to $k=400\,$MeV. With these results we discuss the CP violation in the $\Xi\rightarrow\Lambda\pi$ and $\Omega\rightarrow\Xi\pi$ weak decays. 

\medskip
\ni PACS numbers: 13.75.Gx$\,,$ 13.88.+e 
\end{abstract}

\bigskip

\section{Introduction}

Why should we study pion-hyperon ($\pi Y$) interaction? It is not hard to 
see that, due to their instability, it is not an easy task for an 
experimentalist to make beams of pions and hyperons, let them collide and 
study what happens in such collisions. As far as we know, no experimental 
data on $\pi Y$ interaction are available. In such a situation, is there 
any practical interest, besides academic one, in theoretically studying 
these interactions?

In 1957, Okubo\cite{okubo} observed that the $CP$ violation allows 
$\Sigma$ and $\bar\Sigma$ to have different branching ratios into 
conjugate channels. Pais\cite{pais} extended this proposal also to 
$\Lambda$ and $\bar\Lambda$ decays. In these reactions, the final-state 
strong interaction between the decay products plays a very important role. 
The few studies on $\pi Y$ interactions we could find in the literature 
\cite{nath,datta,kammal} are related to the $\Xi\rightarrow\pi\Lambda$ 
decay, in which an independent estimate of the $\pi \Lambda $ strong 
phase shifts is needed to correctly analyze the data and conclude about 
the CP violation. In these references, however, the results presented 
show some discrepancy among them, especially on $\delta _{S}$, requiring 
a clarification. As for the other interactions, such as $\pi\Sigma$ and 
$\pi \Xi $, within our limited knowledge no study has ever been done. 

Besides, we have a somewhat different motivation for the present study. 
It is by now well known that in high-energy proton-nucleus collisions, 
the inclusively produced hyperons appear usually 
polarized\cite{bunce,ho,mor}. Several models have been proposed to 
explain this phenomenon\cite{ander,dgran,soff,trosh,kubo}, which at 
least qualitatively, or even quantitatively, can account for the 
{\bf hyperon} polarization. However, as 
for the {\bf anti-hyperons} which are generally produced also with 
polarization\cite{ho,mor}, no one of these models are applicable, since 
all of them are based on some leading-particle effect in which the 
incident proton is transformed into a leading hyperon.\footnote{It should 
be mentioned that in a recent paper\cite{anselmino}, a parametrization of 
$\Lambda$ and $\bar\Lambda$ polarization data has been carried out in terms 
of polarizing fragmentation functions.} In \cite{hama}, it 
is proposed that at least part of the polarization is caused by the 
final-state interaction of (anti-)hyperon with the surrounding hot medium 
where it is produced during the collision of the incident objects. This 
mechanism would be the dominant one in the case of anti-hyperon 
polarization, since they cannot be produced as leading particles. 
In \cite{hama}, this idea was put forward within a hydrodynamic model, by 
treating the interaction with the hot medium as given by an optical 
potential, reproducing all the qualitative features of the existing data. 
Evidently, it is desirable that, if possible, more realistic microscopic 
interaction be used instead of purely phenomenological potential with 
fitted parameters. Since pions are dominant in such a hot medium 
mentioned above, the microscopic interactions of our interest would be 
pion-hyperon (or more precisely pion-anti-hyperon) interactions. However, 
except for few results on $\pi\Lambda$, we are not aware of any study on 
these interactions. So the main object of the present work is to study 
the low-energy (with respect to the surrounding medium) pion-hyperon 
interactions, aiming at a later computation of anti-hyperon polarization 
in high-energy hadron-nucleus collisions. 

The plan of presentation is the following. We shall first explain, in the 
next section, the general strategy of treating the pion-hyperon 
interactions. Then, in sections III, IV and V, we apply it, respectively, 
to the $\pi-\Lambda$, $\pi-\Sigma$ and $\pi-\Xi$ cases. Phase shifts are 
calculated and from these the energy dependence of the total cross-section, 
the angular distribution and the polarization for each reaction are 
computed in these sections. Conclusions are drawn in section VI. Basic 
formalism is given in the appendix. 

\section{Strategy for the study of the pion-hyperon interactions}

How could we proceed to study the low-energy $\pi \bar{Y}$ interactions?
First of all, due to the CPT invariance, it is enough to study the 
$\pi Y$ interactions instead of the $\pi \bar{Y}$ ones. For instance, the 
$\bar{Y}$ polarization is obtained from the corresponding one for $Y$, 
just by changing the sign. Next, recall that unlike the $\pi Y$ ones, the 
low-energy $\pi N$ interaction is, for obvious reasons, very well studied 
since a long time. There is a large amount of experimental data, and also 
many models\cite{ols,pearce,bonfinger,gross,goudsmit,ellis} that 
reproduce them pretty well. Here, we shall consider a chiral-invariant 
effective Lagrangian model. In \cite{mane}, such a Lagrangian was written 
in terms of $\pi $, $N$, $\rho$ and $\Delta $ fields as a sum of 

\begin{eqnarray} 
{\cal{L}}_{N\pi N}&=&{\frac{g}{2m}}{[\overline{N}}\gamma_{\mu}\gamma_{5} 
\vec{\tau}N]\cdot\partial^{\mu}\vec{\phi}\ ,\label{pinn} \\ 
&&\nonumber \\
{\cal{L}}_{N\pi\Delta} &=&g_{\Delta}\{{\overline{\Delta}^{\mu}}
[g_{\mu\nu}\!-(Z+{\frac{1}{2}})\gamma _{\mu}\gamma_{\nu}]\vec{M}N\}\cdot
\partial^{\nu}\vec{\phi}  \nonumber \\ 
&&+\ H.c.\ ,\label{pind} \\ 
{\cal{L}}_{N\rho N} &=&{\frac{g_0}{2}}[{\overline{N}}\gamma_{\mu} 
\vec{\tau}N]\cdot\vec{\rho}^{\mu}+{\frac{g_0}{2}}[{\overline{N}}
({\frac{\mu_p-\mu_n}{4m}})i\sigma_{\mu\nu}\vec{\tau}N]  \nonumber \\ 
&&\cdot\,(\partial^\mu\vec{\rho^{\nu}}-\partial^{\nu}\vec{\rho^\mu}) 
\ ,\label{rhonn} \\
{\cal{L}}_{\pi\rho\pi}&=&g_{0}\,\vec{\rho}_{\mu}\cdot(\vec{\phi}\times 
\partial^{\mu}\vec{\phi})-{\frac{g_0}{4m_{\rho}^2}}
(\partial_{\mu}\vec{\rho}_{\nu}-\partial_{\nu}\vec{\rho}_{\mu})\nonumber\\ 
&&\cdot\,(\partial^{\mu}\vec{\phi}\times\partial^{\nu}\vec{\phi})
\ ,\label{rhopp}
\end{eqnarray}

\ni where $N$, $\Delta$, $\vec\phi$, $\vec\rho$ are the nucleon, delta, pion 
and rho fields with masses $m$, $m_\Delta$, $m_\pi$, and $m_\rho\,$, 
respectively, $\mu_p$ and $\mu_n$ are the proton and neutron magnetic
moments, $\vec M$ and $\vec\tau$ are the isospin matrices and $Z$ is a 
parameter representing the possibility of the off-shell-$\Delta$ having 
spin 1/2. In addition, it also included a $\sigma$ term as a correction and 
parametrized it in a way we will show below.

Now, since $\pi\Lambda$, $\pi\Sigma$ and $\pi\Xi$ systems are similar
to $\pi N$, we can make an analogy and use the same prescription explained
above, adapting it appropriately. The $\Delta (1232)$ resonance plays a
central role in the low-energy $\pi N$ interaction. Its contribution
dominates the total cross section of $\pi ^{+}p$ ($T=3/2$) process and is
also important to the other isospin channels. The lowest energy hyperon
resonances and their main decay modes are quite well known, so it is
possible to use these resonances replacing $\Delta (1232)$. 
As for the coupling constants, they can be estimated from the resonance 
widths \cite{kammal}.   

Another detail we have to take into account is the unitarization of the
amplitudes. In an effective model like the one we are considering, the
amplitudes we directly obtain are real, consequently violate the
unitarity of the $S$ matrix. So, if we want something more than simple 
cross section, some procedure is required to unitarize the amplitudes. As 
is often done in effective models\cite{bonfinger,goudsmit,ellis,andrea}, 
and will be explained in detail in the next section, we will do this by 
reinterpreting the calculated amplitudes as elements of reaction matrix $K$. 

Now we are ready to calculate all the phase shifts and then the total cross 
sections, angular distributions and polarizations. Because we are interested 
in low energies ($k\le$ .4 GeV), we will limit ourselves to the $S$ and $P$ 
waves, which are generally enough for our purpose.

\section{Pion-Lambda Interaction}

The $\pi\Lambda$ interaction is the simplest case. Since $\Lambda$ has 
isospin 0, the scattering amplitude $T_{\pi\Lambda}$ has the general 
form 
\begin{equation}
T_{\pi\Lambda }^{ba}=\overline{u}(\vec{p}\,\prime)[A(k,\theta)+ 
{\frac{(\not{\!}k\,+\!\not{\!}k\prime)}{2}}B(k,\theta)]\delta_{ba}
u(\vec{p})\ \ , 
\end{equation}
where $p_{\mu}$ and $p_{\mu}^{\prime}$ are the initial and final 
4-momenta of $\Lambda$ in the center of mass frame, $k_{\mu}$ and 
$k_{\mu}^{\prime}$ are those of the pion, and $\theta$ the scattering 
angle. Indices $a$ and $b$ indicate the initial and final isospin states 
of the pion. We show in Fig.~1 the relevant diagrams, where we have omitted 
the crossed diagrams, although included in the calculations. We consider 
only the first resonance $\Sigma^{\ast}(1385)$, because we are interested in 
the low-energy ($k\leq $0.4 GeV) behavior. The $\rho$ exchange term is absent 
in the $\pi\Lambda$ case, because due to the isospin it does not couple to 
$\Lambda$. To computing the first two of these diagrams, the Lagrangians 
(\ref{pinn}) and (\ref{pind}) have been adapted to 

\begin{eqnarray}
{\cal {L}}_{\Lambda\pi\Sigma}\,
&=& \frac{g_{\Lambda\pi\Sigma}}{2m_\Lambda}\lbrack {\overline\Sigma_\alpha}\gamma_\mu\gamma_{5}\Lambda\rbrack\partial^\mu 
\phi_\alpha + H.c.\ , \label{LpS}\\
{\cal {L}}_{\Lambda\pi\Sigma^*}\!&=& g_{\Lambda\pi\Sigma^*} \lbrace 
{\overline{\Sigma^*}_\alpha^\mu}\lbrack g_{\mu\nu}\!-(Z+{\frac{1}{2}}) 
\gamma_\mu\gamma_\nu\rbrack\Lambda\rbrace\partial^\nu\phi_\alpha+ H.c.\,, 
\nonumber \\
\end{eqnarray}
by replacing the nucleon by $\Lambda$ or $\Sigma$, and $\Delta$ by 
$\Sigma^{\ast}$ and performing appropriate sums over isotopic spin indices. 

\begin{figure}[hbtp] 
\centerline{
\epsfxsize=8.cm
\epsffile{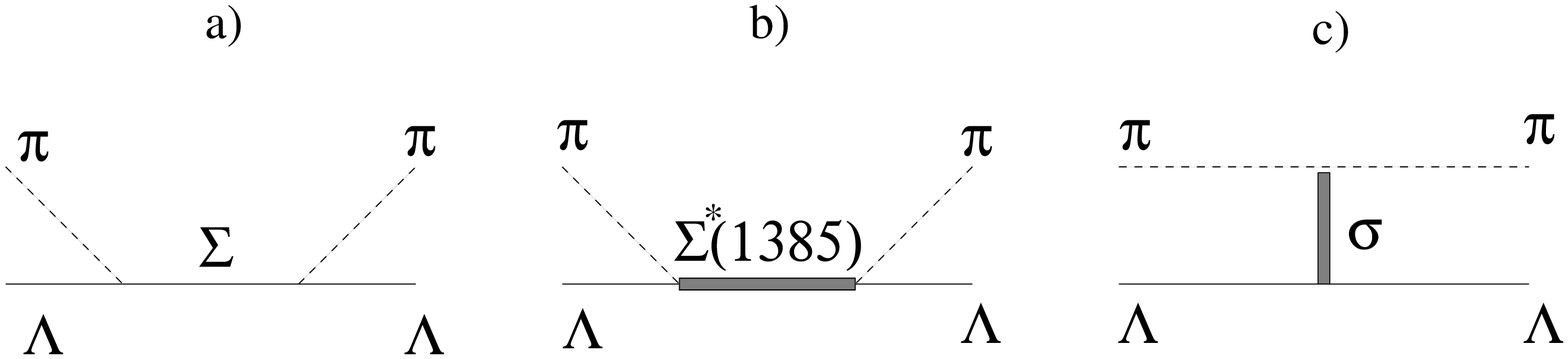}}
\centerline{\small{Fig. 1: Diagrams for $\pi\Lambda$ Interaction}}
\end{figure}

The contributions of the diagram (1a) to the amplitudes are  
\begin{eqnarray}
A_\Sigma &=& \frac{g_{\Lambda\pi\Sigma}^2}{4\,m^2_\Lambda}
 (m_\Lambda+m_\Sigma) \lbrace {\frac{s-m_\Lambda^2}{s-m_\Sigma^2}} 
 +{\frac{u-m_\Lambda^2}{u-m_\Sigma^2}}\rbrace\ , \nonumber \\
B_\Sigma &=& \frac{g_{\Lambda\pi\Sigma}^2}{4\,m^2_\Lambda} 
 \lbrace{\frac{m_\Lambda^2-s-2m_\Lambda(m_\Lambda+m_\Sigma)}{s-m_\Sigma^2}} 
 \nonumber \\
&&+{\frac{2m_\Lambda(m_\Lambda+m_\Sigma)+ u-m_\Lambda^2}{u-m_\Sigma^2}} 
 \rbrace \ \ .
\end{eqnarray}

The diagram (1b) gives  
\begin{eqnarray}
A_{\Sigma^*} &=& {\frac{g_{\Lambda\pi\Sigma^*}^2}{3m_\Lambda}}\lbrace 
{\frac{\nu_r}{\nu_r^2-\nu^2}}\hat A-{\frac{m_\Lambda^2+m_\Lambda m_{\Sigma^*}}{m_{\Sigma^*}^2}}  \nonumber \\ 
&&\hspace{1.cm}\times(2m_{\Sigma^*}^2+m_\Lambda m_{\Sigma^*}
-m_\Lambda^2+2m_\pi^2)  \nonumber \\
&&+{\frac{4m_\Lambda}{m_{\Sigma^*}^2}}\lbrack(m_\Lambda\!+
m_{\Sigma^*})Z\!+\!(2m_{\Sigma^*}\!+m_\Lambda)Z^2\rbrack k.k^{\prime}
\rbrace\ , \nonumber \\
B_{\Sigma^*} &=& {\frac{g_{\Lambda\pi\Sigma^*}^2}{3m_\Lambda}} 
\lbrace{\frac{\nu}{\nu_r^2-\nu^2}}\hat B - {\frac{8m_\Lambda^2\nu Z^2} 
{m_{\Sigma^*}^2}}\rbrace\ ,
\end{eqnarray}
where $\nu$ and $\nu_r$ are defined in the Appendix and 
\begin{eqnarray}
\hat A&=&{\frac{(m_{\Sigma^*}+m_\Lambda)^2-m_\pi^2}{2m_{\Sigma^*}^2}}
\lbrack 2m_{\Sigma^*}^3-2m_\Lambda^3-2m_\Lambda m_{\Sigma^*}^2  \nonumber \\
&&-2m_\Lambda^2m_{\Sigma^*}\!\!+\!m_\pi^2(2m_\Lambda\!\!-\!m_{\Sigma^*})
\rbrack+{\frac{3}{2}}(m_\Lambda\!\!+\!m_{\Sigma^*})t\ ,  \nonumber \\
\hat B&=&{\frac{1}{2m_{\Sigma^*}^2}}\lbrack
(m_{\Sigma^*}^2-m_\Lambda^2)^2-2m_\pi^2(m_{\Sigma^*}^2+m_\Lambda)^2
+m_\pi^4\rbrack  \nonumber \\
&&+ {\frac{3}{2}}\,t\ .
\end{eqnarray}

As for the diagram (1c), we only parametrize the amplitudes as done in 
\cite{ols}
\begin{eqnarray}
A_{\sigma } &=&a+bt\ , \nonumber \\
B_{\sigma } &=&0\ ,\label{sigma} 
\end{eqnarray}
where $a=1.05\,m_\pi^{-1}$ and $b=-0.80\,m_\pi^{-3}$ are constants (we use 
the same values of \cite{mane} for $\pi N$). The scattering matrix will 
then have the form 
\beq
M_{ba}={\frac{T_{ba}}{8\pi\sqrt{s}}}=f_{1}+{\frac{(\vec{\sigma}.\vec{k}) 
(\vec{\sigma}.\vec{k}\prime)}{kk\prime}}f_{2} 
\eeq
and we can make the partial wave decomposition with 
\beq
a_{l\pm }={\frac{1}{2}}\int_{-1}^{1}[P_{l}(x)f_{1}(x)+P_{l\pm 1}(x)f_{2}(x)
]\ \ . 
\eeq

The amplitudes $a_{l\pm}\,$, calculated in a tree-level approximation, are 
real and, so, the corresponding $S$ matrix is not unitary. In order to 
unitarize these amplitudes, we reinterpret them as elements of $K$ matrix and 
write 
\beq
a_{l\pm }^{U}={\frac{a_{l\pm }}{1-ik\ a_{l\pm }}} \ \ . 
\eeq
The phase-shifts are then computed as 
\beq
\delta_{l\pm} = {\rm tg}^{-1}(k\ a_{l\pm}) \ \ . 
\eeq

The parameters we use are $m_{\Lambda}=1.115\,$GeV, $m_{\Sigma}=1.192\,$GeV, 
$m_{\Sigma^{\ast}}=1.385\,$GeV, $m_\pi=0.139\,$GeV \cite{data}, 
$g_{\Lambda\pi\Sigma}=11.7\,$\cite{martin,pilk} and $Z=-0.5\,$\cite{mane}. 
The only parameter that is missing is $g_{\Lambda\pi{\Sigma^{\ast}}}$. 
As mentioned before, we estimate it from the resonance width. Namely, by 
comparing the $\delta_{P3}$ phase-shift in the resonance region with the 
relativistic Breit-Wigner expression \cite{perl}, 
\beq
\delta_{l\pm}={\rm tg}^{-1}[{\frac{\Gamma_0({\frac{k}{k_0}})^{2l}} 
{2(m_r-\sqrt{s})}}]\ \ , 
\eeq 
where $k_0$ is the center-of-mass momentum at the peak of 
$\Sigma^{\ast}(1385)$, that is $0.207\,$GeV. The value obtained in this way 
is $g_{\Lambda\pi\Sigma^{\ast}}=9.38\,$GeV$^{-1}$, which we will use here.  

In Fig.~2, we show the calculated phase-shifts as functions of the 
center-of-mass momentum $k\,$. We also show there the $k$ dependence of 
the total elastic cross section, the angular distribution and the $\Lambda$ 
polarization as function of $x=\cos\theta\,$, for $k=100, 200, 300\ 
{\rm and}\ 400\,$MeV.

\vspace{-1.cm} 
\begin{figure}[h]
\hspace{-.2cm}
\epsfxsize=118.mm
\epsffile{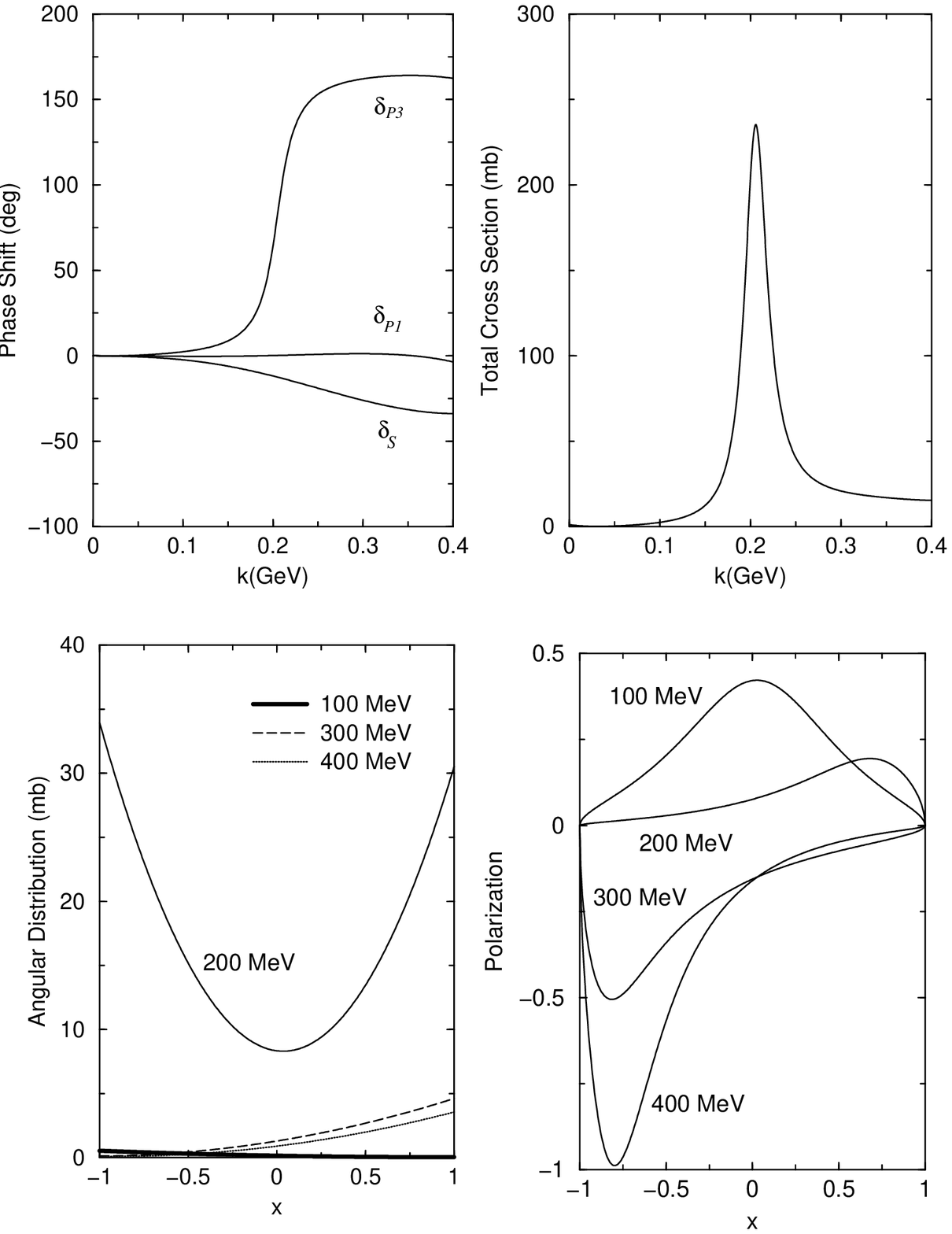}

\vspace{-4.5cm} 
\centerline{\small{Fig. 2: $\pi\Lambda$ Scattering}}
\end{figure}

As we can see, the $\Sigma^*(1385)$ contribution dominates the total elastic
cross section in the low energy region (quite similar to $\pi^+p$
scattering). As for the polarization, it begins positive at lower energies 
and then becomes negative above $k\sim k_0$.

\section{Pion-Sigma Interaction}

In the case of $\pi\Sigma$ interaction, both $\pi$ and $\Sigma$ have isospin 
1, so the compound system can have isospin 2,1 or 0. For this reason, the 
scattering amplitude is somewhat more complex in this case and has the 
following general form 
\begin{eqnarray}
T_{\alpha\gamma,\beta\delta}&=&
 \langle\pi_\gamma\Sigma_\delta|T|\pi_\alpha\Sigma_\beta\rangle\nonumber \\ 
&=&\overline{u}(\vec{p}\,\prime)\lbrace\lbrack A+  
 {\frac{(\not\!k\,+\!\not\!k^{\prime})}{2}}A^{\prime}\rbrack 
 \delta_{\alpha\beta}\delta_{\gamma\delta} \nonumber \\ 
&&+\lbrack B+{\frac{(\not\!k\,+\!\not\!k^{\prime})}{2}}B^{\prime} 
 \rbrack\delta_{\alpha\gamma}\delta_{\beta\delta}  \nonumber \\ 
&&+\lbrack C+{\frac{(\not\!k\,+\!\not\!k^{\prime})}{2}}C^{\prime} 
 \rbrack\delta_{\alpha\delta}\delta_{\beta\gamma}\rbrace u(\vec p)\ ,
\end{eqnarray}
where $\alpha,\,\beta,\,\gamma$ and $\delta$ are isospin indices.  
Decomposing this amplitude into the $i$-th. isospin states of the system 
($P_i$ are the projection operators),  
we have 
\begin{eqnarray} 
T_{\alpha\gamma,\beta\delta}&=&\overline{u}(\vec{p}\,\prime)\lbrace \lbrack
 A_0+{\frac{(\not\!k\,+\!\not\!k^{\prime})}{2}}B_0\rbrack P_0  \nonumber \\
&&\!+\lbrack A_1\!+\!{\frac{(\not\!k+\!\!\not\!k^{\prime})}{2}}B_1 \rbrack
 P_1\!+\!\lbrack A_2\!+\! {\frac{(\not\!k+\!\!\not\!k^{\prime})}{2}}
 B_2\rbrack P_2\rbrace u(\vec p)  \nonumber \\
&=&\overline{u}(\vec{p}\,\prime)\lbrace{\frac{1}{3}}\lbrack A_0+ 
 {\frac{(\not\!k\,+\!\not\!k^{\prime})}{2}}B_0\rbrack\delta_{\alpha\beta} 
 \delta_{\gamma\delta}  \nonumber \\ 
&&+{\frac{1}{2}}\lbrack A_1+{\frac{(\not\!k\,+\!\not\!k^{\prime})}{2}}
 B_1\rbrack\lbrack\delta_{\alpha\gamma}\delta_{\beta\delta}-
 \delta_{\alpha\delta}\delta_{\beta\gamma}\rbrack  \nonumber \\
&&+{\frac{1}{6}}\lbrack A_2+{\frac{(\not\!k\,+\!\not\!k^{\prime})}{2}}
 B_2\rbrack  \nonumber \\
&&\ \ \times\lbrack3\delta_{\alpha\gamma}\delta_{\beta\delta}+
 3\delta_{\alpha\delta}\delta_{\beta\gamma}-\delta_{\alpha\beta}
 \delta_{\gamma\delta}\rbrack\rbrace u(\vec p)\ .
\end{eqnarray}
Comparing (17) and (18) we obtain 
\beq
\cases{A_0=3A+B+C\cr 
       A_1= B-C\cr 
       A_2=B+C\cr} \ \ , \ \ 
\cases{B_0=3A'+B'+C'\cr 
       B_1= B'-C'\cr 
       B_2=B'+C'\cr} \ \ . 
\eeq
These are the relations that determine all the amplitudes projected on 
isospin states. 

The interaction Lagrangians are given by (\ref{rhopp}), (\ref{LpS}) and 
\begin{eqnarray}
{\cal {L}}_{\Sigma\pi\Sigma}&=&{\frac{g_{\Sigma\pi\Sigma}}{2m_\Sigma}} 
\lbrack{\overline\Sigma} \gamma_\mu\gamma_5\,\vec t\,\Sigma\rbrack\cdot%
\partial^\mu\vec\phi\ , \\
{\cal {L}}_{\Sigma\rho\Sigma}&=&{\frac{g_0}{2}} \lbrack{\overline\Sigma}\gamma_\mu\vec t\,\Sigma \rbrack\cdot\vec\rho^\mu+{\frac{g_0}{2}}\lbrack{\overline\Sigma} ({\frac{\mu_{\Sigma^0}-\mu_{\Sigma^-}}{4m_\Sigma}})\,i\sigma_{\mu\nu}\, 
\vec t\,\Sigma\rbrack  \nonumber \\
&&\cdot(\partial^\mu\vec{\rho^\nu}-\partial^\nu\vec{\rho^\mu})\ ,
\end{eqnarray}
where the isospin combination matrix $\vec t$ obeys 
\beq
\langle\beta|\vec t|\alpha\rangle=-i\epsilon_{\beta\alpha c}\hat e_c\ . 
\eeq

\begin{figure}[hbtp]
\centerline{
\epsfxsize=8.cm
\epsffile{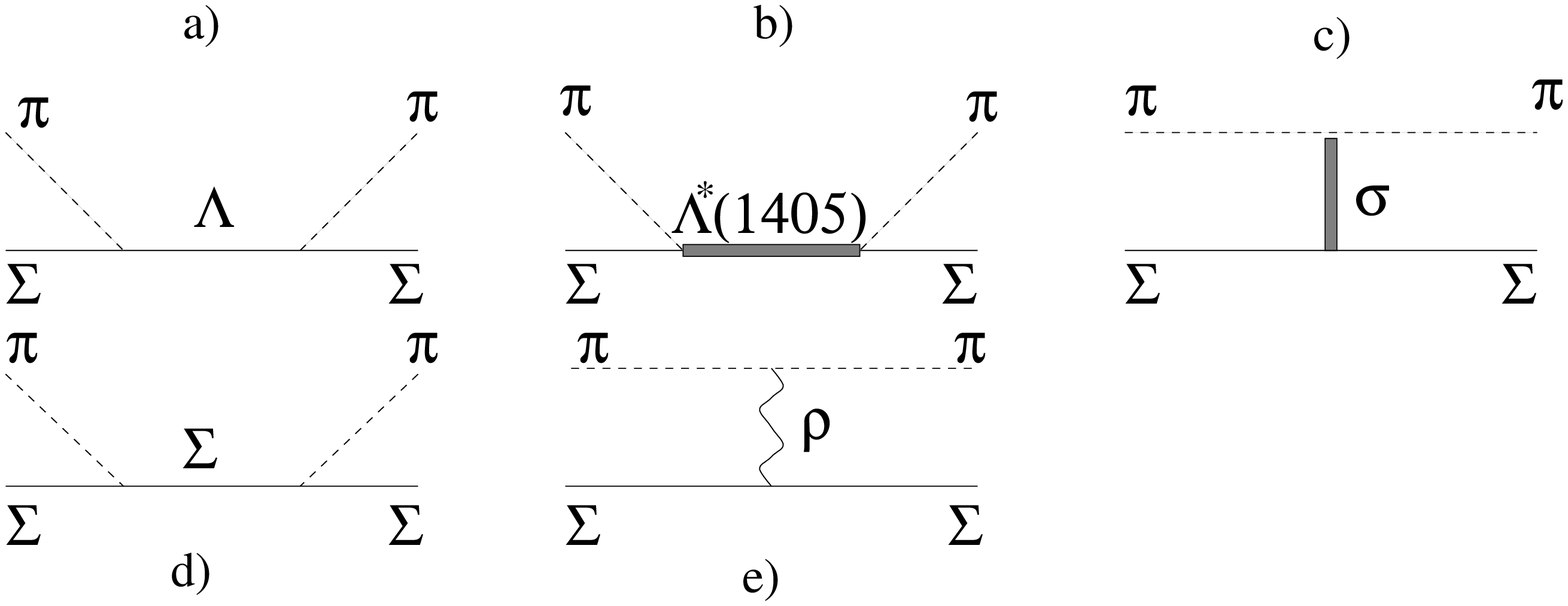}}
\centerline{\small{Fig. 3: Diagrams to $\pi\Sigma$ Interaction}}
\end{figure}

Figure 3 shows the diagrams we consider for the $\pi\Sigma$ interactions. 
$\Sigma^*(1385)$ also couples to $\pi\Sigma$, but its decay branching ratio 
back to the $\pi\Sigma$ channel is only 11\%$\,$. So, we will neglect it. 

The amplitudes corresponding to the diagram a), Fig.~3, are 
\begin{eqnarray}
A_\Lambda &=& \frac{g_{\Lambda\pi\Sigma}^2}{4\,m_\Sigma^2} 
 (m_\Sigma+m_\Lambda) {\frac{s-m_\Sigma^2}{s-m_\Lambda^2}}\ , \nonumber \\
B_\Lambda &=& 0\ ,  \nonumber \\
C_\Lambda &=& \frac{g_{\Lambda\pi\Sigma}^2}{4\,m_\Sigma^2} 
 (m_\Sigma+m_\Lambda) {\frac{u-m_\Sigma^2}{u-m_\Lambda^2}}\ ,  \nonumber \\
A_\Lambda^{\prime}&=& \frac{g_{\Lambda\pi\Sigma}^2}{4\,m_\Sigma^2}  
 {\frac{m_\Sigma^2-s-2m_\Sigma(m_\Sigma+m_\Lambda)}{s-m_\Lambda^2}}\ ,  
 \nonumber \\
B_\Lambda^{\prime}&=& 0\ ,  \nonumber \\
C_\Lambda^{\prime}&=& \frac{g_{\Lambda\pi\Sigma}^2}{4\,m_\Sigma^2}  
 {\frac{2m_\Sigma(m_\Sigma+m_\Lambda)+u-m_\Sigma^2}{u-m_\Lambda^2}}\ . 
\end{eqnarray}
The contributions of the diagram b) with intermediate $\Lambda^{\ast}(1405)$ 
are similar. We must only change the coupling constant and replace the mass 
$m_\Lambda$ by $m_{\Lambda^{\ast}}$. 

In the case of the intermediate $\Sigma$, diagram d), we have 
\begin{eqnarray}
A_\Sigma &=&-{\frac{g_{\Sigma\pi\Sigma}^2}{2m_\Sigma}}\ ,  \nonumber \\
B_\Sigma &=& {\frac{g_{\Sigma\pi\Sigma}^2}{ m_\Sigma}}\ ,  \nonumber \\
C_\Sigma &=&-{\frac{g_{\Sigma\pi\Sigma}^2}{2m_\Sigma}}\ ,  \nonumber \\
A_\Sigma^{\prime}&=&-{\frac{g_{\Sigma\pi\Sigma}^2}{4m_\Sigma^2}} 
 -{\frac{g_{\Sigma\pi\Sigma}^2}{2m_\Sigma}}{\frac{1}{\nu_0-\nu}}\ ,  
 \nonumber \\
B_\Sigma^{\prime}&=& {\frac{g_{\Sigma\pi\Sigma}^2}{ m_\Sigma}} 
 {\frac{\nu}{\nu_0^2-\nu^2}}\ ,  \nonumber \\
C_\Sigma^{\prime}&=&{\frac{g_{\Sigma\pi\Sigma}^2}{4m_\Sigma^2}} +{\frac{%
g_{\Sigma\pi\Sigma}^2}{2m_\Sigma}}{\frac{1}{\nu_0+\nu}}\ .
\end{eqnarray}

The $\rho$ exchange amplitude, diagram e), has the form 
\beq
T_\rho=\overline{u}(\vec p')\lbrack A_\rho\!+ {\frac{(\not\!k+\!\!\not\!k')}{%
2}}B_\rho\rbrack\lbrack  \delta_{\alpha\beta}\delta_{\gamma\delta}- 
\delta_{\alpha\delta}\delta_{\beta\gamma}\rbrack u(\vec p)\ , 
\eeq
so 
\beq
\cases{A_0=2A_\rho\cr 
       A_1=A_\rho\cr 
       A_2=-A_\rho\cr} \ \ , \ \ 
\cases{B_0=2B_\rho\cr 
       B_1=B_\rho\cr 
       B_2=-B_\rho\cr} \ \ , 
\eeq
with 
\begin{eqnarray}
&A_\rho&=-{\frac{g_0^2}{m_\rho^2}}\,(\mu_{\Sigma^0}-\mu_{\Sigma^-})
 \nu\, {\frac{1 -t/4m_\rho^2}{1 - t/ m_\rho^2}}\ ,  \nonumber \\
&B_\rho&={\frac{g_0^2}{m_\rho^2}}\,(1+\mu_{\Sigma^0}-\mu_{\Sigma^-})\, 
 {\frac{1 -t/4m_\rho^2}{1 - t/m_\rho^2}} \ .
\end{eqnarray}

Finally, the $\sigma$-term has been parametrized in the same way as for 
$\pi\Lambda$, by using eqs. (\ref{sigma}), with the same parameters. 
In addition to the parameters used in the $\pi\Lambda$ case, we use here 
$m_{\Lambda ^{\ast}}=1.406$ GeV ,
$m_{\rho}$=.769 GeV, $\mu_{\Sigma^{0}}=.649$, $\mu_{\Sigma^-}=-0.16$ and 
$g_{\Sigma\pi\Sigma }$=6.7\cite{pilk}.
The coupling constant $g_{\Sigma\pi\Lambda^{\ast}}$ is not known, but we can 
proceed in the same way as we have done before, comparing the calculated 
amplitudes with the Breit-Wigner expression. The best fit is obtained with 
$g_{\Sigma\pi\Lambda^{\ast}}= 8.74\,$GeV$^{-1}$. 

We show in Fig.~4 the phase shifts calculated as explained above. It is also 
shown the energy dependence of the cross section $\sigma_t$ for each channel 
described below. 

\vspace{-1.5cm} 
\begin{figure}[h]
\hspace{-.2cm}
\epsfxsize=115.mm
\epsffile{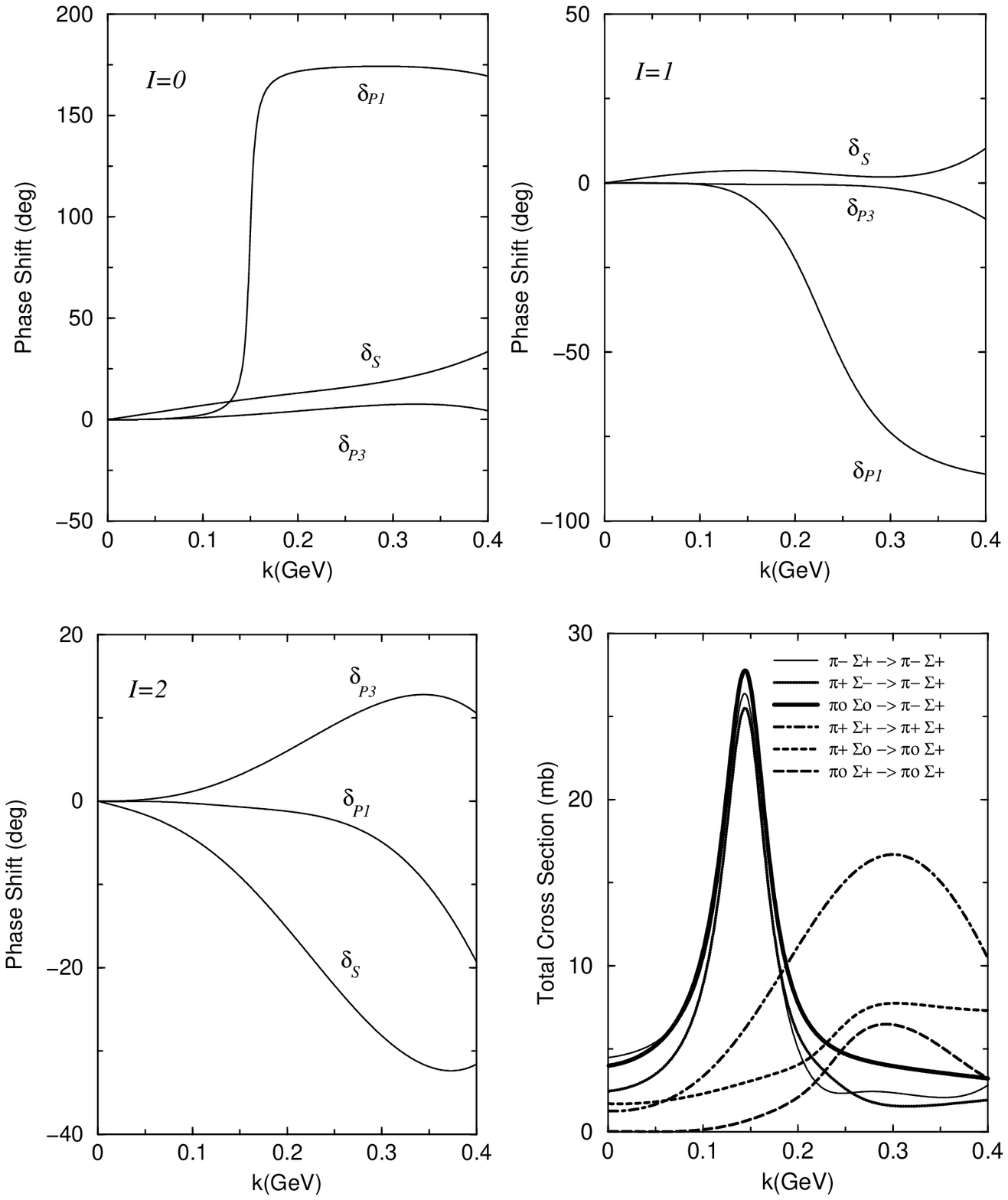}

\vspace{-4.5cm} 
\centerline{\small{Fig. 4: Phase-shifts and the energy dependence of 
$\sigma_t$ for $\pi\Sigma$}} 
\ni{\small{interactions}}
\end{figure}

Using the isospin formalism we calculate the elastic, as well as the charge 
exchange, amplitudes as 
\begin{eqnarray}
 \langle\pi^+\Sigma^+|T|\pi^+\Sigma^+\rangle 
  &=&\langle\pi^-\Sigma^-|T|\pi^-\Sigma^-\rangle 
   = T_2  \nonumber \\ 
 \langle\pi^+\Sigma^0\,|T|\pi^+\Sigma^0\,\rangle 
  &=&\langle\pi^-\Sigma^0\,|T|\pi^-\Sigma^0\,\rangle  
   =\langle\pi^0\Sigma^+|T|\pi^0\Sigma^+\rangle  \nonumber \\
  &=&\langle\pi^0\,\Sigma^-|T|\pi^0\,\Sigma^-\rangle   
   ={\frac{T_2}{2}}+{\frac{T_1}{2}}  \nonumber \\
 \langle\pi^0\,\Sigma^0\,|T|\pi^0\,\Sigma^0\,\rangle 
  &=&{\frac{2\,T_2}{3}}+{\frac{T_0}{3}}  \nonumber \\
 \langle\pi^+\Sigma^-|T|\pi^+\Sigma^-\rangle 
  &=&\langle\pi^-\Sigma^+|T|\pi^-\Sigma^+\rangle 
   ={\frac{T_2}{6}}+{\frac{T_1}{2}}+{\frac{T_0}{3}}  \nonumber \\ 
 \langle\pi^-\Sigma^+\!|T|\pi^+\Sigma^-\!\rangle 
  &=&\langle\pi^+\Sigma^-\!|T|\pi\!^-\Sigma^+\rangle 
   ={\frac{T_2}{6}}-{\frac{T_1}{2}}+{\frac{T_0}{3}}  \nonumber \\ 
 \langle\pi^+\Sigma^0|T|\pi^0\Sigma^+\rangle 
  &=&\langle\pi^-\Sigma^0|T|\pi^0\Sigma^-\rangle 
   =\langle\pi^0\Sigma^+|T|\pi^+\Sigma^0\rangle  \nonumber \\ 
  &=&\langle\pi^0\Sigma^-|T|\pi^-\Sigma^0\rangle 
   ={\frac{T_2}{2}}-{\frac{T_1}{2}}  \nonumber \\ 
 \langle\pi^+\Sigma^-|T|\pi^0\Sigma^0\rangle 
  &=&\langle\pi^-\Sigma^+|T|\pi^0\Sigma^0\rangle 
   =\langle\pi^0\Sigma^0|T|\pi^+\Sigma^-\rangle  \nonumber \\ 
  &=&\langle\pi^0\Sigma^0|T|\pi^-\Sigma^+\rangle 
   ={\frac{T_2}{3}}-{\frac{T_0}{3}}\ . 
\end{eqnarray} 

\vspace*{-1.5cm} 
\begin{figure}[h]
\epsfxsize=145.mm
\hspace*{-0.2cm}
\epsffile{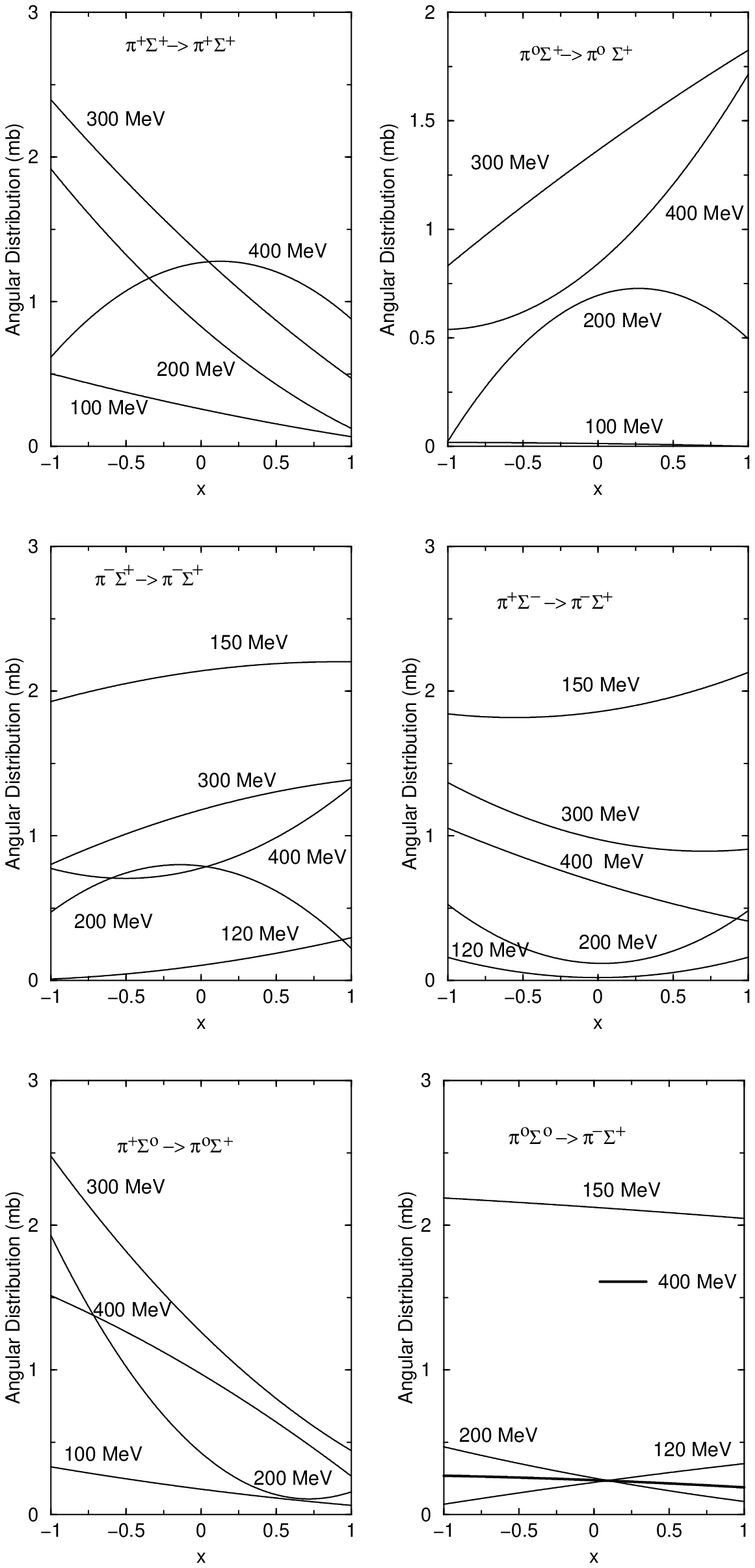}

\vspace*{-2.0cm} 
\centerline{\small{Fig. 5: Angular distributions of $\Sigma^+$}}
\end{figure} 

\smallskip

With these amplitudes, we can calculate $\sigma_t$, $d\sigma/d\Omega$ and 
$P$ as functions of $k$ and $x=\cos\,\theta$ for each channel. The results 
for $\Sigma^+$ in the final state are shown in Figs.~$4-6$.

\vspace*{-1.2cm} 
\begin{figure}[hbtp]
\epsfxsize=140.mm 
\hspace*{-0.2cm} 
\epsffile{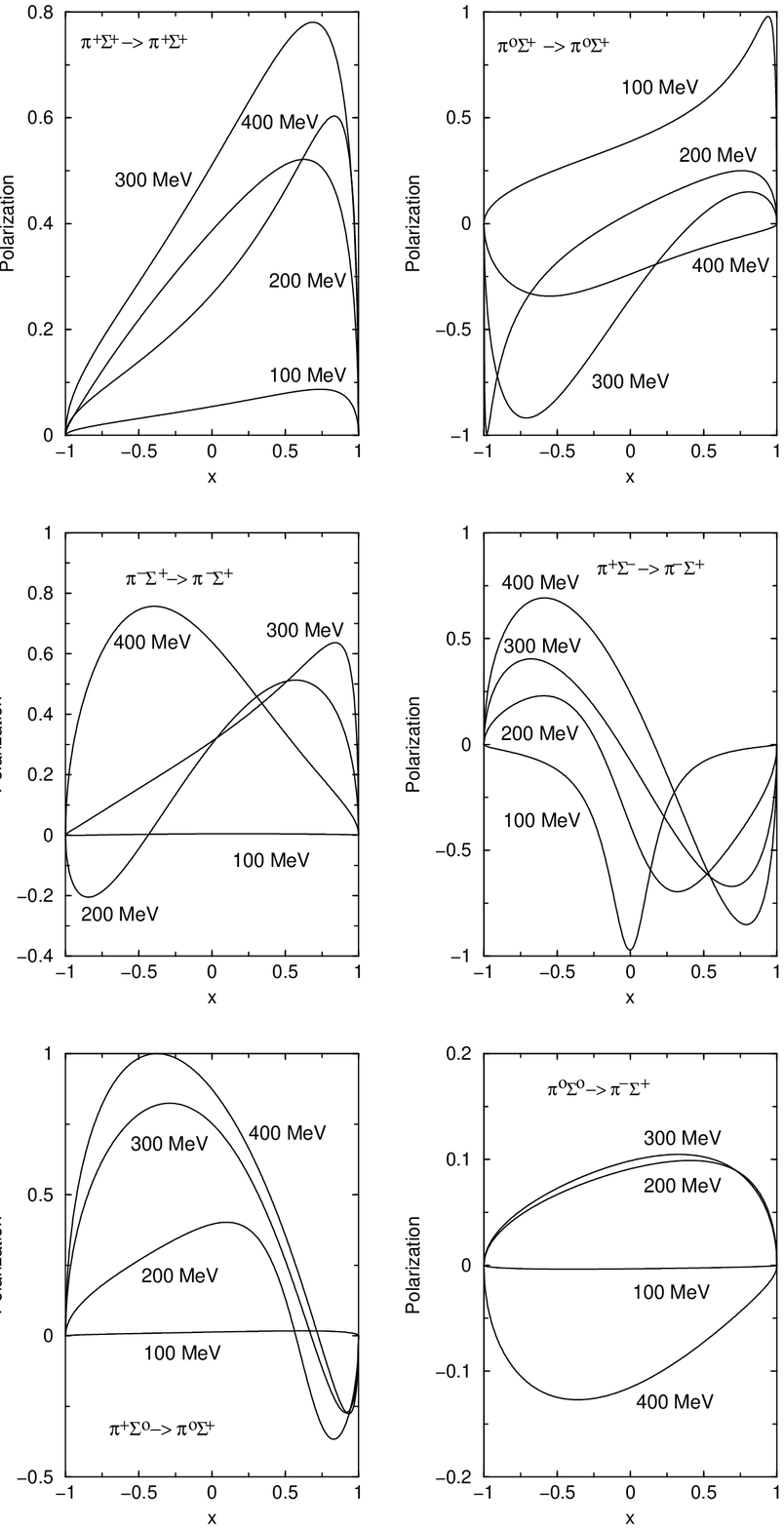}

\vspace*{-2.0cm} 
\centerline{\small{Fig. 6: Polarization of $\Sigma^+$}}
\end{figure}

We can see in Fig.~4 that, although the first resonance is important in 
$\pi\Sigma$ interactions, it is not as much as in the $\pi^{+}p$ or 
$\pi\Lambda$ scatterings. The peak in the $\Lambda^{\ast }(1405)$-mass 
region is not so high (less than 30 mb) and it appears in the $I=0$ state 
($\pi^{-}\Sigma^{+}$). Remark that the other reactions ($I=1$ and 
especially $I=2$) have comparable total cross sections. 

Before passing to the next section, it is worth-while making the following 
remarks. Even in the tree-level calculation and in the low-energy 
($k\lesssim 0.4$ GeV) region that we are considering here, there could occur 
the exchange reactions $\pi\Lambda\rightleftharpoons\pi\Sigma\,$. The 
possible diagrams for these are similar to Fig.~1 b) and Figs.~3 b) and e), 
with one of $\Lambda$ ($\Sigma$) replaced by $\Sigma$ ($\Lambda$). However, 
the contributions of these reactions are small compared with the elastic ones we examined in this paper. First, as mentioned before and could be 
seen in Figs.~2 and 4, the direct-resonances dominate over all the other 
processes, which appear as corrections to the former. They don't change 
much the cross sections, however are necessary to produce polarization. Now, 
$\pi\Sigma\rightarrow\pi\Lambda\,$, which is given by the $\Sigma^*$ term 
together with $\rho$ exchange one, is much smaller than 
$\pi\Lambda\rightarrow\pi\Lambda\,$, because the branching ratio of  
$\Sigma^*$ decay is 
$(\Sigma^*\rightarrow\pi\Sigma)/(\Sigma^*\rightarrow\pi\Lambda)\sim0.16\,$
\cite{data}. As for $\pi\Lambda\rightarrow\pi\Sigma\,$ compared with 
$\pi\Sigma\rightarrow\pi\Sigma\,$, as mentioned above first we have 
$\sigma(\pi\Lambda\rightarrow\pi\Sigma)
/\sigma(\pi\Lambda\rightarrow\pi\Lambda)\sim0.16$ for each possible channel. 
Now, from Figs. 2 and 4, each $\pi\Lambda\rightarrow\pi\Lambda$ channel, 
compared with the sum of the three prominent $\pi\Sigma\rightarrow\pi\Sigma$ 
channels, gives $\sigma(\pi\Lambda\rightarrow\pi\Lambda)/ \sum\sigma(\pi\Sigma\rightarrow\pi\Sigma)\sim0.60$ on the average in the 
resonance region. So, we estimate that the overall 
$\pi\Lambda\rightarrow\pi\Sigma$ contribution is less than 20\% of 
$\pi\Sigma\rightarrow\pi\Sigma$ examined here. 

\section{${\bf \protect\pi\Xi}$ Interaction}

This case is very similar to the $\pi N$ scattering, because $\Xi$ 
has isospin 1/2 (as the nucleon) and the main difference is that the 
resonance of interest $\Xi^*(1533)$ has isospin $I$=1/2 (instead of $I$=3/2 
as $\Delta(1232)$). Then, the scattering amplitude $T_{\pi\Xi}^{ba}$ has the 
general form 
\begin{eqnarray}
T_{\pi\Xi}^{ba}&=&\overline{u}(\vec{p}\,\prime)\lbrace\lbrack A^+ +
 {\frac{(\not\!k+\!\!\not\!k^{\prime})}{2}}B^+\rbrack\delta_{ba}\nonumber \\
 &&+\lbrack A^- +{\frac{(\not\!k+\!\!\not\!k^{\prime})}{2}}B^-\rbrack
i\epsilon_{bac}\tau^c\rbrace u(\vec p)\ .
\end{eqnarray}
The contributing diagrams are in Fig.~7 and the Lagrangians are almost the 
same as in the case of $\pi N$ scattering, eqs. (\ref{pinn}-\ref{rhopp}), 
where we must replace the $N$ field by $\Xi$ field, and $\Delta(1232)$ by 
$\Xi^*(1533)$. The latter implies a substitution of the isospin matrix 
$\vec M$ by $\vec \tau$. Consequently, $A_{\Xi^*}^{\pm}$ and 
$B_{\Xi^*}^{\pm}$ have different structures as compared with 
$A_{\Delta}^{\pm}$ and $B_{\Delta}^{\pm}$ of $\pi N$ case, whereas all the 
other $A^{\pm}$ and $B^{\pm}$ remain the same, with appropriate parameter changes.   
\begin{figure}[hbtp]
\centerline{
\epsfxsize=8.cm
\epsffile{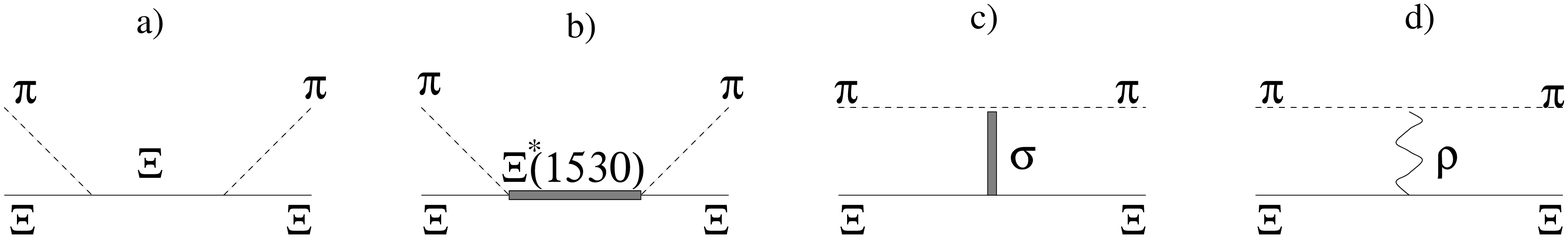}}
\centerline{\small{Fig. 7: Diagrams to $\pi\Xi$ Interaction}}
\end{figure}

So, by computing the Feynman diagram a) in Fig.~7, we obtain 
\begin{eqnarray}
A_\Xi^+&=& {\frac{g_{\Xi\pi\Xi}^2}{ m_\Xi}}\ ,  \nonumber \\
A_\Xi^-&=& 0\ ,  \nonumber \\
B_\Xi^+&=& {\frac{g_{\Xi\pi\Xi}^2}{ m_\Xi}}{\frac{\nu}{\nu_0^2-\nu^2}} \ , 
 \nonumber \\
B_\Xi^-&=&-{\frac{g_{\Xi\pi\Xi}^2}{2m_\Xi^2}}
 -{\frac{g_{\Xi\pi\Xi}^2}{m_\Xi}} {\frac{\nu_0}{\nu_0^2-\nu^2}}\ .
\end{eqnarray}
The $\rho$ exchange, diagram d), gives 
\begin{eqnarray}
A_\rho^+&=& B_\rho^+ = 0\ ,  \nonumber \\
A_\rho^-&=&-{\frac{g_0^2}{m_\rho^2}}(\mu_{\Xi^0}-\mu_{\Xi^-})\nu {\frac{1-{t/4m_\rho^2}}{1-{t/m_\rho^2}}}\ ,  \nonumber \\
B_\rho^-&=& {\frac{g_0^2}{m_\rho^2}}(1+\mu_{\Xi^0}-\mu_{\Xi^-}) 
{\frac{1-{t/4m_\rho^2}}{1-{t/m_\rho^2}}}\ .
\end{eqnarray}
The contributions from diagram b) with intermediate $\Xi^*(1533)$ are

\begin{eqnarray}
A_{\Xi^*}^+&=&{\frac{g_{\Xi\pi\Xi^*}^2}{3m_\Xi}}\lbrace {\frac{\nu_r} 
 {\nu_r^2-\nu^2}}\hat A-{\frac{m_\Xi^2+m_\Xi m_{\Xi^*}}{m_{\Xi^*}^2}} 
 \nonumber \\
&&\hspace{1.cm}\times(2m_{\Xi^*}^2+m_\Xi m_{\Xi^*}-m_\Xi^2+2m_\pi^2) 
 \nonumber \\
&&+{\frac{4m_\Xi}{m_{\Xi^*}^2}}\lbrack (m_\Xi+m_{\Xi^*})Z+(2m_{\Xi^*} 
 +m_\Xi)Z^2\rbrack k.k^{\prime}\rbrace\ ,  \nonumber \\
A_{\Xi^*}^-&=&{\frac{g_{\Xi\pi\Xi^*}^2}{3m_\Xi}}\lbrace {\frac{\nu}
 {\nu_r^2-\nu^2}}\hat A+{\frac{8m_\Xi^2\nu}{m_{\Xi^*}^2}}  \nonumber \\ 
&&\hspace{1.cm}\times\lbrack(m_\Xi+m_{\Xi^*})Z+(2m_{\Xi^*}+m_\Xi)Z^2
 \rbrack\rbrace\ ,  \nonumber \\
B_{\Xi^*}^+&=&{\frac{g_{\Xi\pi\Xi^*}^2}{3m_\Xi}}\lbrace {\frac{\nu} 
 {\nu_r^2-\nu^2}}\hat B-{\frac{8m_\Xi^2\nu Z^2}{m_{\Xi^*}^2}}\rbrace \ , 
 \nonumber \\
B_{\Xi^*}^-&=&{\frac{g_{\Xi\pi\Xi^*}^2}{3m_\Xi}}\lbrace{\frac{\nu_r} 
 {\nu_r^2-\nu^2}}\hat B -m_\Xi{\frac{(m_\Xi+m_{\Xi^*})^2}{m_{\Xi^*}^2}} 
 -{\frac{4m_\Xi Z^2}{m_{\Xi^*}^2}}k.k^{\prime}  \nonumber \\ 
&&-{\frac{4m_\Xi}{m_{\Xi^*}^2}}\lbrack(2m_\Xi^2+2m_\Xi m_{\Xi^*}-2m_\pi^2)Z 
 \nonumber \\
&&\hspace{1.cm}+(2m_\Xi^2+4m_\Xi m_{\Xi^*})Z^2\rbrack\rbrace\ .
\end{eqnarray}
where 
\begin{eqnarray}
\hat A&=&{\frac{(m_{\Xi^*}+m_\Xi)^2-m_\pi^2}{2m_{\Xi^*}^2}}\lbrack
2m_{\Xi^*}^3-2m_\Xi^3-2m_\Xi m_{\Xi^*}^2  \nonumber \\
&&-2m_\Xi^2m_{\Xi^*}+m_\pi^2(2m_\Xi-m_{\Xi^*})\rbrack+ {\frac{3}{2}}%
(m_\Xi+m_{\Xi^*})t\ ,  \nonumber \\
\hat B&=&{\frac{1}{2m_{\Xi^*}^2}}\lbrack(m_{\Xi^*}^2\!-m_\Xi^2)^2\!
-2m_\pi^2(m_{\Xi^*}^2\!+m_\Xi)^2\!+m_\pi^4\rbrack+{\frac{3}{2}}t\ . 
\nonumber \\
\end{eqnarray} 

The parameters used are $m_\Xi$=1.318 GeV, $m_{\Xi^*}$=1.533 GeV, 
$\mu_{\Xi^0}=-1.25$, $\mu_{\Xi^-}=0.349$ and $g_{\Xi\pi\Xi}=4$. As in the 
previous cases, we determined the $\Xi\pi\Xi^*$ coupling constant by using 
the Breit-Wigner formula and got the value 
$g_{\Xi\pi\Xi^*}=4.54\,$GeV$^{-1}$. We display in Fig.~8 the calculated 
phase shifts to the isospin 1/2 and 3/2 states. 

\vspace*{-.8cm} 
\begin{figure}[hb]
\epsfxsize=118.mm 
\hspace*{-0.4cm} 
\epsffile{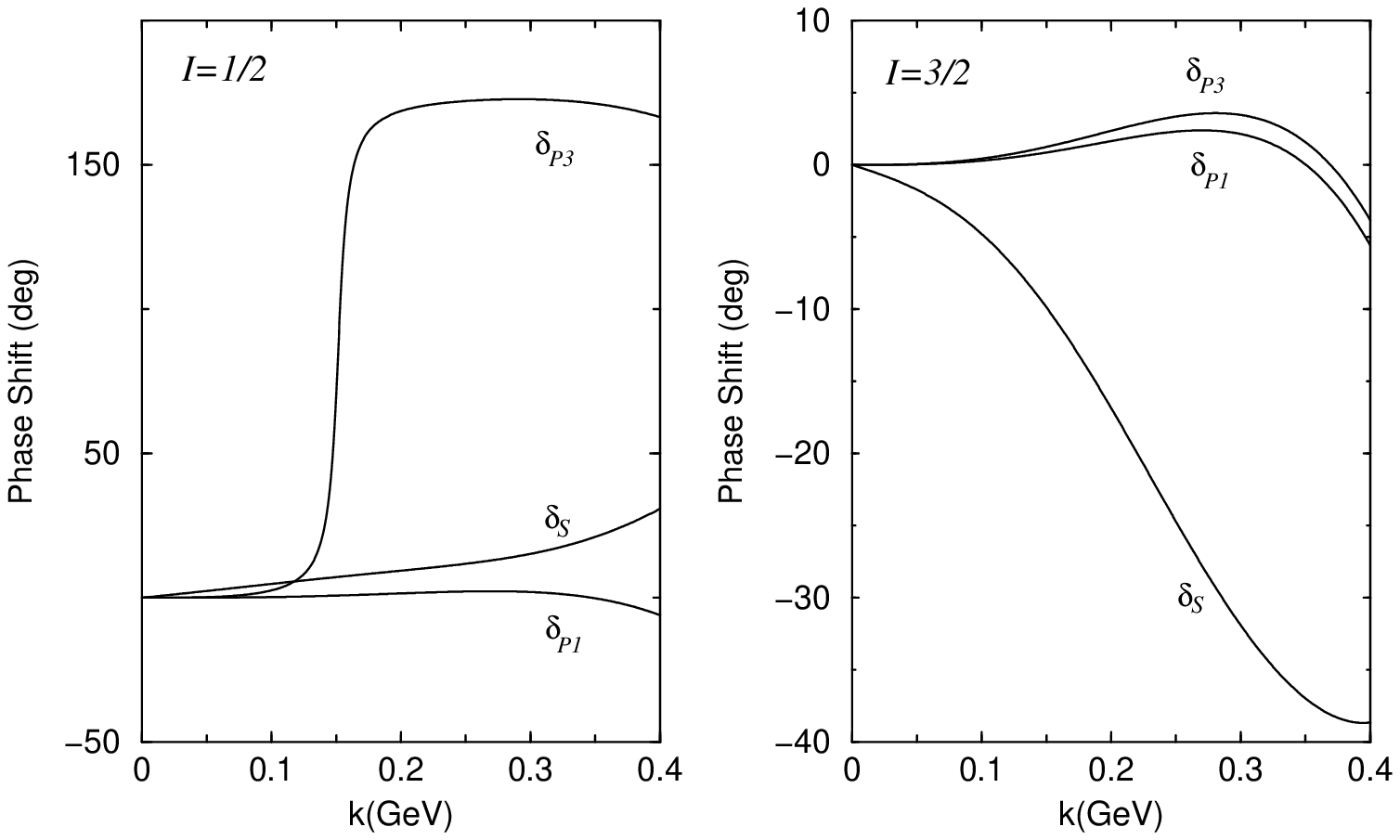}

\vspace*{-10.0cm} 
\centerline{\small{Fig. 8: Phase shifts for $\pi\Xi$ interaction}} 
\end{figure}

\medskip

We can now obtain the matrix elements for each elastic and 
charge-exchange channel as 

\begin{eqnarray}
&\langle\pi^+\Xi^0|T|\pi^+\Xi^0\rangle&=
\langle\pi^-\Xi^-|T|\pi^-\Xi^-\rangle = T_{\frac{3}{2}}\ , \nonumber \\
&\langle\pi^+\Xi^-|T|\pi^+\Xi^-\rangle&=\langle\pi^-\Xi^0|T|\pi^-\Xi^0 
\rangle = {\frac{1}{3}}T_{\frac{3}{2}}+{\frac{2}{3}}T_{\frac{1}{2}}\ ,  \nonumber \\
&\langle\pi^0\Xi^-|T|\pi^0\Xi^-\rangle&=\langle\pi^0\Xi^0|T|\pi^0\Xi^0 \rangle = {\frac{2}{3}}T_{\frac{3}{2}}+{\frac{1}{3}}T_{\frac{1}{2}}\ , \nonumber \\
&\langle\pi^0\Xi^-|T|\pi^-\Xi^0\rangle&=
\langle\pi^+\Xi^-|T|\pi^0\Xi^0\rangle = {\frac{\sqrt{2}}{3}}T_{\frac{3}{2}} -{\frac{\sqrt{2}}{3}}T_{\frac{1}{2}}\ ,  \nonumber \\
&\langle\pi^-\Xi^0|T|\pi^0\Xi^-\rangle&=
\langle\pi^-\Xi^+|T|\pi^0\Xi^0\rangle = {\frac{\sqrt{2}}{3}}T_{\frac{3}{2}} -{\frac{\sqrt{2}}{3}}T_{\frac{1}{2}}\ . \nonumber \\
\end{eqnarray}

\begin{figure}[hbtp]
\epsfxsize=44.mm 
\hspace*{1.8cm} 
\epsffile{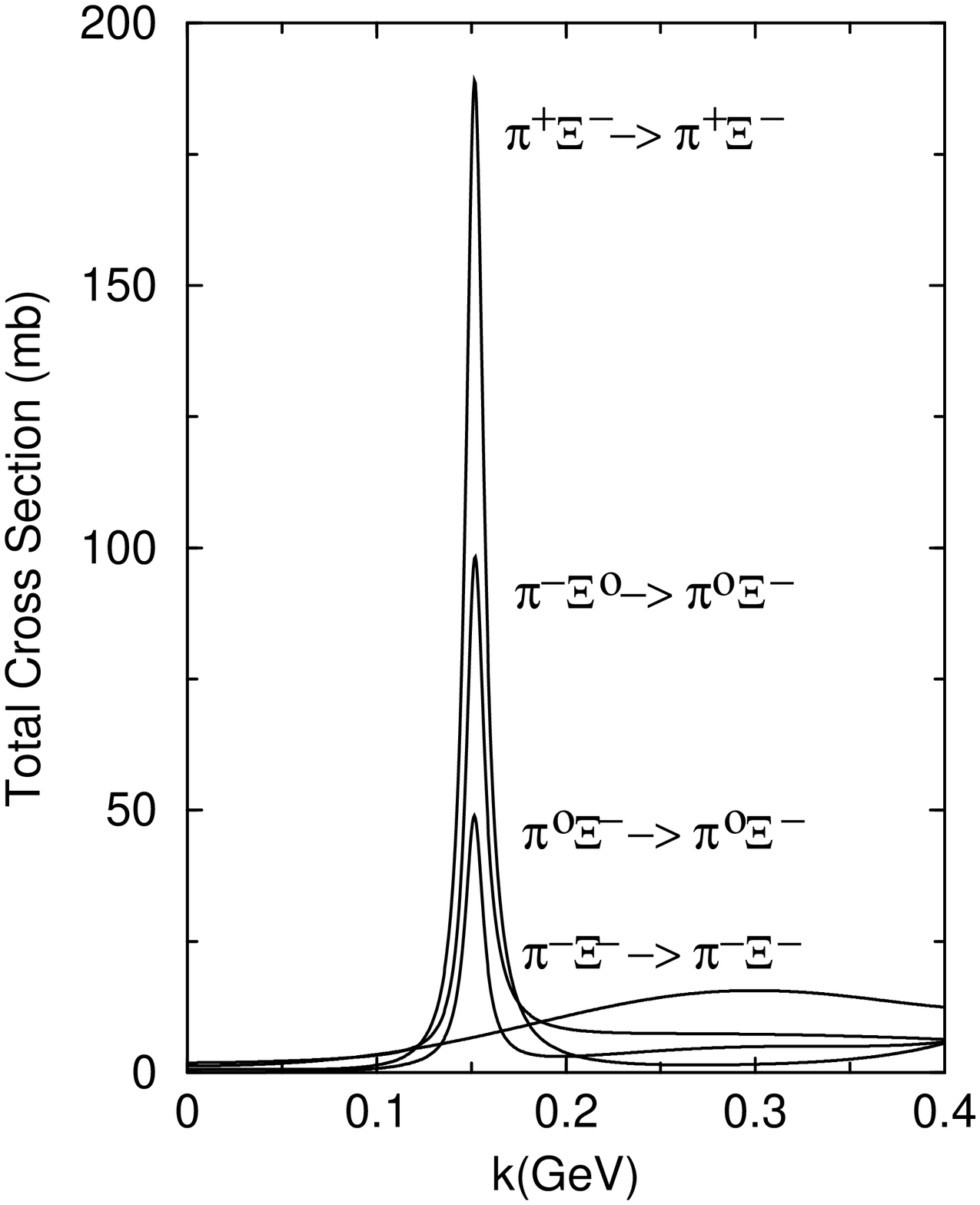} 

\vspace{.2cm} 
\centerline{\small{Fig. 9: Total cross sections for $\pi\Xi$ interaction}}
\end{figure} 

\vspace*{-2.cm} 
\begin{figure}[hbtp]
\epsfxsize=160.mm
\epsffile{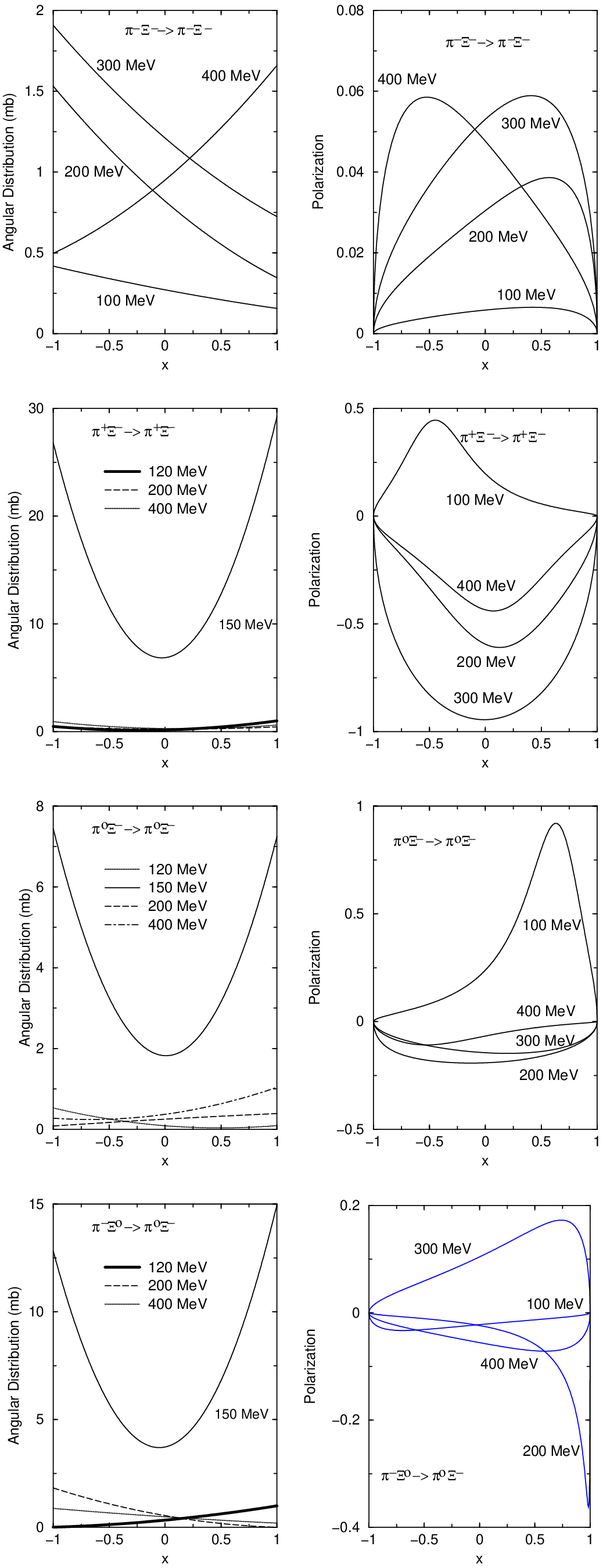} 
\centerline{\small{Fig. 10: $d\sigma/d\Omega$ and Polarizations for 
$\Xi^-$ production}}
\end{figure}

We show, in Fig.~9, the integrated cross sections, with $\Xi^-$ in the final 
state, obtained with these matrix elements. We can see that in this 
case the $\Xi(1533)$ resonance contribution is very important and it 
dominates three of the reactions. Figure 10 presents the angular 
distributions and polarizations for the same reactions. 

\section{Discussion of the Results}

In the preceding sections, by making a close analogy with the well 
established $\pi N$ case, we have calculated the $S$- and $P$-wave phase 
shifts for $\pi\Lambda$, $\pi\Sigma$ and $\pi\Xi$ interactions. Then obtained both the integrated and differential cross sections and 
polarizations for all the elastic and charge-exchange processes. Let us 
now discuss these results in connection with the two applications we 
mentioned in the introduction. 

The first application refers to the study of the CP violation. One of the 
ways to verify this violation is to observe the hyperon weak decays, 
$\Lambda\rightarrow\pi N$, $\Sigma\rightarrow\pi N$, 
$\Xi\rightarrow\pi\Lambda$ and $\Omega\rightarrow\pi\Xi$. In such a study, 
we need an independent estimate of the strong-interaction phase shifts in 
the final state. 


For $\Lambda$- and $\Sigma$-decays, a large amount of data are available 
on the strong interaction phase shifts, since $\pi N$ scatterings are very 
well studied. In the $\Xi$ decay, there are some estimates of the $S$- and 
$P1$-wave phase shifts for $\pi\Lambda$ system. However, the reported 
results are conflicting with each other. Whereas the authors of \cite{nath} 
give $\delta_{S}=-18.7^{o}$ and $\delta _{P1}=-2.7^{o}$, in \cite{kammal}, 
they tell that $\delta _{S}=1.2^{o}$ and $\delta _{P1}=-1.7^{o}$ and, as 
for the Ref. \cite{datta}, $\delta_{S}$ is between $-1.3^{o}$ and 
$0.1^{o}$ and $\delta_{P1}$ between $-0.4^{o}$ and $-3.0^{o}$. 
In our calculation, with the $\sigma$ term included, we obtained 
$\delta_{P1}=-0.36^{o}$ and $\delta _{S}=-4.69^{o}$ at the $\Lambda$-mass 
value, that gives $\delta_{S}-\delta_{P1}\sim-4.3^{o}$, that is still small. 
One should remark that to really fit the phase shifts in $\pi N$ scattering, 
especially $\delta_{S}$, it is necessary to include other contributions as 
the diffractive \cite{ols} or the contact \cite{ellis} terms with correct 
parameters. So it is possible that some correction is needed in the results 
we have obtained here.

In this paper, we have also calculated the $\pi\Xi$ phase shifts. So, it 
is possible to get some information about the CP violation in the 
$\Omega\rightarrow\pi\Xi$ decay, too. $\Omega$ has $J^p={3\over2}^+$, so 
the phase shifts we need are $\delta_{P3}^I$ and $\delta_{D3}^I$. 
Calculating the asymmetry parameter $A$ in the same way as in 
\cite{kammal}, the approximate expression reads  
\begin{eqnarray}
A&=&-{\rm tan}(\delta_{P3}^{1\over 2}-\delta_{D3}^{1\over 2})
{\rm tan}(\phi_{P3}^{1\over 2}-\phi_{D3}^{1\over 2})\nonumber \\ 
 &\sim &-{\rm tan}(\delta_{P3}^{1\over 2})
{\rm tan}(\phi_{P3}^{1\over 2}-\phi_{D3}^{1\over 2})
\   \  .
\end{eqnarray} 

\ni 
At the $\Omega$ mass value, $\delta_{D3}^{1\over 2}=0.21^o$ (computed with 
the same diagrams used in sec. 5) and 
$\delta_{P3}^{1\over 2}=173.04^o=180^o-6.96^o$. So the strong 
interaction effect in the asymmetry parameter will appear as 
tan($-7.17^o$) that is a value close to that obtained in the $\Xi$ decay. 
So we do not expect that, in the study of CP violation in hyperon weak 
decays, $\Omega\rightarrow\pi\Xi$ is much useful.  

The other application we mentioned in the introduction, and which was the 
main motivation of this work, is the inclusive (anti-)hyperon polarization in high-energy collisions. As explained there, the anti-hyperon polarization 
cannot be understood in terms of the usual models 
\cite{ander,dgran,soff,trosh,kubo}, because all of them are based on the 
leading-particle effect and an anti-hyperon cannot be a leading particle. 
In \cite{hama} it has been proposed that anti-hyperons are polarized when 
interacting with the surrounding particles, which are predominantly pions, 
that make the environment where they are produced. So the anti-hyperon 
polarization would appear as an average effect of the low energy 
$\pi\overline{Y}$ interaction. 
It is clear that, generally speaking, such an average procedure washes out 
any existent asymmetry, so that no polarization would appear as 
a consequence. This is true if we look at the the central region of the 
collision. However, the polarization data are obtained in very forward 
directions where the asymmetry could be preserved. 
Such calculations will be reported elsewhere\cite{celso}, but just 
observing the results of the preceding sections we can draw some 
conclusions. The $\Lambda$ polarization, as seen in Fig.~2, is positive 
below 100 Mev and then changes the sign, so we expect that, on averaging, 
the most part will be canceled out, implying that the polarization of 
${\overline{\Lambda }\sim 0}$. 
As seen in Figs.~9 and 10, the $\Xi^{-}$ polarization is negative and very 
large in the channels where the cross section is large, whereas the 
$\Sigma^{-}$ polarization is positive in most of the cases, Figs.~6. As we 
can see, the hyperon polarization is different in each case, and seems to 
be consistent with the experimental data for the anti-hyperons\cite{ho,mor}. 
Remark that the polarization sign changes under charge conjugation. 

\vspace{.6cm}

\ni {\bf Acknowledgments}

\bigskip

We would like to thank M.R. Robilotta for the discussions about the hadron 
interactions and comments on the manuscript. This work was partially 
supported by CNPq and FAPESP (contract nos. 98/02249-4 and 00/04422-7).

\bigskip

\appendix
\renewcommand{\theequation}{\Alph{section}\arabic{equation}}

\setcounter{equation}{0}

\section{Basic Formalism}

In this paper $p$ and $p^{\prime}$ are the initial and final hyperon
4-momenta, $k$ and $k^{\prime}$ are the initial and final pion 4-momenta, so
the Mandelstam variables are 
\begin{eqnarray}
s &=& (p+k)^2=(p^{\prime}+k^{\prime})^2 \\
t &=& (p-p^{\prime})^2=(k-k^{\prime})^2 \\
u &=& (p^{\prime}-k)^2=(p-k^{\prime})^2 \ \ .
\end{eqnarray}
With these variables, we can define 
\begin{eqnarray}
\nu &=& {\frac{s-u}{4m}} \\
\nu_0 &=& {\frac{2m_\pi^2-t}{4m}} \\
\nu_r &=& {\frac{m_r^2-m^2-k.k^{\prime}}{2m}} \ \ ,
\end{eqnarray}
where $m$, $m_r$ and $m_\pi$ are, respectively, the hyperon mass, the 
resonance mass and the pion mass. The scattering amplitude for an isospin 
$I$ state is 
\beq
T_I=\overline{u}(\vec p')\lbrace \lbrack A^I + {\frac{(\not\!k + \not\!k')}{2}}B^I\rbrack\rbrace u(\vec p)\ ,  
\eeq

\ni where $A_I$ and $B_I$ are calculated using the Feynman diagrams. So the 
scattering matrix is 
\beq
M_I^{ba} = {\frac{T_I^{ba}}{8\pi\sqrt{s}}} = f_I(\theta) + \vec\sigma.\hat n
g_I(\theta) = f_1^I + {\frac{(\vec\sigma .\vec k' )(\vec\sigma .\vec k)}{kk'}}f_2^I \ \ , 
\eeq
with 
\begin{eqnarray}
& &f_1^I(\theta) = {\frac{(E+m)}{8\pi\sqrt{s}}} \lbrack A_I + (\sqrt{s}%
-m)B_I\rbrack\ \ , \\
& & f_2^I(\theta) = {\frac{(E-m)}{8\pi\sqrt{s}}} \lbrack -A_I + (\sqrt{s}%
+m)B_I\rbrack \ \ ,
\end{eqnarray}
where $E$ is the hyperon energy. The partial-wave decomposition is done
with 
\beq
a_{l\pm} = {\frac{1}{2}}\int_{-1}^{1}\lbrack P_l(x)f_1(x) + P_{l\pm
1}(x)f_2(x)  \rbrack dx \ \ . 
\eeq

In our calculation (tree level) $a_{l\pm}$ is real. With the unitarization, 
as explained in Section III, we obtain 
\beq
 a_{l\pm}^U = {\frac{1}{2ik}}\lbrack e^{2i\delta_{l\pm}}-1\rbrack 
 = \frac{e^{i\delta_{l\pm}}}{k}\,{\rm sen}(\delta_{l\pm})
 \rightarrow a_{l\pm} \ \ . 
\eeq
These complex amplitudes are used to calculate

\begin{eqnarray}
&& f(\theta)=\sum_{l=0}^{\infty}\lbrack (l+1)a_{l+}+la_{l-}\rbrack\,P_l(x)
 \ , \\
&& g(\theta)=i\sum_{l=1}^{\infty}\lbrack a_{l+}-a_{l-}\rbrack\,P_l^{(1)}(x)
 \ . 
\end{eqnarray}
We have, then, in the center-of-mass frame, 

\begin{eqnarray}
& {\frac{d\sigma}{d\Omega}} & = |f|^2 + |g|^2 \ , \\
& \vec P & =-2{\frac{{\rm Im}(f^*g)}{|f|^2 + |g|^2}}\hat n \ , \\
& \sigma_t & = 4\pi \sum_l\lbrack (l+1)|a_{l+}|^2 + l|a_{l-}|^2\rbrack \ .
\end{eqnarray}

\end{document}